\documentclass[12pt]{iopart}
\usepackage{iopams}
\usepackage{graphicx}
\usepackage{bm}
\usepackage{url}
\begin{document}

\title[Finite wavelength cloaking]{Finite wavelength cloaking by plasmonic resonance}

\author{N-A P Nicorovici$^1$, R C McPhedran$^1$, S Enoch$^2$ and G Tayeb$^2$}

\address{$^1$ CUDOS, School of Physics, University of Sydney, NSW 2006, Australia}
\address{$^2$ Institut Fresnel, CNRS, Aix-Marseille Universit\'{e}, 13013 Marseille, France}
\begin{abstract}
We consider cloaking by a coated cylindrical system using plasmonic
resonance, and extend previous quasistatic treatments to include the
effect of finite wavelength. We show that a probe cylinder can still
be cloaked at finite wavelengths, but the cloaking cylinder develops
a non-zero scattering cross-section. We show that this latter effect
is dominated by a monopole term in the case of an ideal (lossless)
cloaking material, and by a dipole term in the case of a realistic
(lossy) material. It can be reduced but not eliminated by variations of
geometric or dielectric parameters of the cloaking cylinder.

\end{abstract}

\pacs{78.20.-e, 41.20.-q, 42.25.Bs}
\submitto{\NJP}
\maketitle

\section{Introduction}

There is much current interest in the possibility of cloaking or hiding objects
from scrutiny by electromagnetic waves. At least three techniques have been proposed
to achieve this:
one avoids detection by surrounding the target body with a metamaterial shell which guides
light around the central cavity \cite{pendry,ulf},
the second again relies on a metamaterial \cite{cc1,cc2,engheta,mnrp,mn,usoptexp},
which this time cloaks by resonance an external region,
while the third uses a structured metamaterial which provides cloaking by in effect
folding space back upon itself \cite{ulf2006,kildishev}. The technical challenges of making
such systems in practice are enormous, but it should be realized that these and other
proposals for cloaking offer complementary characteristics, which implies that work on
a range of them is valuable.

\begin{table}[h]
\caption{\label{tab1}Comparison of three cloaking methods.}
\begin{indented}
\item[]\begin{tabular}{llll}
\hline
&&& \\
{\bf Mechanism} & Refraction & Reaction & Unfolding \\
&&& \\ \hline &&& \\
{\bf Region} & Internal & External & External \\
&&& \\ \hline &&& \\
{\bf Structure} & Metamaterial & Metamaterial & Metamaterial \\
 & Shell: $\varepsilon$, $\mu$ $\ge 0$ & homogeneous & Shell: $\varepsilon$, $\mu$ $\ge 0$ \\
 & vary with position & $\varepsilon_\mathrm{s} + \varepsilon_\mathrm{m} = 0$ & vary with position \\
 &&& \\ \hline &&& \\
{\bf Equations} & 2D, 3D Maxwell & 2D quasistatics & 2D, 3D quasistatics \\
&&& \\ \hline &&& \\
{\bf Experiment} & Yes & No & No \\
&&& \\ \hline &&& \\
{\bf Problems} & Bandwidth, & Bandwidth, & Bandwidth, \\
 & structuring shell, & achieving $\varepsilon_\mathrm{s}$, & structuring shell, \\
 & energy dissipation. & scale size $\ll$ $\lambda$. & scale size $\ll$ $\lambda$. \\
 &&& \\ \hline
\end{tabular}
\end{indented}
\end{table}

We present a number of salient characteristics of the three methods in Table~1.
Cloaking by refraction requires a structured metamaterial shell to divert light around a cavity
in which the object to be hidden is placed (internal cloaking). It is designed using
a full solution of Maxwell equations in two or three dimensions, and there has been
an experimental demonstration of this mechanism in the former case \cite{schurig}.
The second and third methods offer the complementary feature of concealing a body
in a region close to, but outside, the cloaking system (external cloaking).
In the case of cloaking by reaction, the object is concealed by virtue of a
plasmonic resonance, which requires the material in the cylindrical cloaking shell
to have a dielectric constant close to the negative of the dielectric constants in the
core and matrix regions surrounding it. It has been studied to this point mainly in two
dimensions. The third method is the most recent, and has features in common with each
of the other two. It achieves cloaking by using a spatially varying dielectric constant
and magnetic permeability, designed using the same principles of transformation optics
at a basis of refractive cloaking. The goal here, however, is to in effect fold space
back upon itself, and, loosely speaking, to hide the object within the enfolding.

It is our purpose here to study one problematic aspect of cloaking by reaction,
which is implicit in two previous papers \cite{mn,BL}, but which has not been studied
systematically. The problem does not arise for cloaking by refraction or unfolding,
by virtue of their different mechanism of operation, which ensures that both
the cloaking system and the cloaked object are hidden to an equal degree from electromagnetic
probes. As we shall see, this is not necessarily for cloaking by reaction, where it is quite
possible for  a larger cloaking system to successfully cloak a small object, but
to be itself quite visible. Using a visual analogy from the animal kingdom, we refer to
this as "Ostrich Effect": {\em the large object hides the small object, but the
large object does not hide itself}. The possibility of the Ostrich Effect, was signaled
in a paper by Milton, Brian and Willis \cite{mbw}: "Besides invisibility there is what
we call cloaking where the surrounding material does not have to be carefully adapted to
suit the object to be made invisible. The cloaking device may be invisible or visible,
although obviously the former is more interesting."

We take the viewpoint here that for many purposes, the Ostrich Effect will be undesirable,
and so we provide examples of the effect, explore its underlying physics, and quantify the
circumstances under which it is greatly reduced. In Sec.~II we present two figures
taken from simulations showing the quenching of the dipole moment of a probe cylinder
in the vicinity of a cylindrical shell with realistic values of the complex
dielectric constant. These figures illustrate clearly the Ostrich Effect.
In the next Section we consider a coated cylinder interacting with an incident plane wave
and solve this scattering problem in closed form. We also take the long wavelengths limit
of the formulation, in order to exhibit the transition from dynamics to qualitative.
In Sec.~IV we use the scattering cross section of the coated cylinder to show the
counterintuitive result that a small imaginary part of $\varepsilon_\mathrm{s}$ actually
benefits cloaking at finite wavelengths, since it makes dipole rather than monopole
terms dominant in the scattering cross section. This is in keeping with the results
of Hao-Yuan She \etal \cite{ojfm} but not with those reported by
Min Yan \etal \cite{qiu}.

\section{Description of cloaking numerical simulations}

Let us consider a two-dimensional physical system comprising a coated cylinder
centred about the origin of coordinates and a probe (solid) cylinder on the y-axis.
Both cylinders are perpendicular to the $xy$-plane.
The shell and core radii of the coated cylinder are, respectively, $r_\mathrm{c} = 20$nm,
$r_\mathrm{s} = 65$nm, while the radius of the probe cylinder is $a = 5$nm.
Also, the core and shell relative permittivities are $\varepsilon_\mathrm{c} = 1$,
$\varepsilon_\mathrm{s} = -1 + 0.1\,\mathrm{i}$. The relative permittivity of the probe cylinder is
$\varepsilon = \varepsilon_\mathrm{s}$ and the relative permittivity of the matrix is $\varepsilon_\mathrm{m} = 1$.
All the components are non-magnetic so that the relative permeabilities are
$\mu_\mathrm{c} = \mu_\mathrm{s}=\mu=\mu_\mathrm{m} = 1$, where $\mu$ is the permeability of the probe cylinder.

\begin{figure}[h]
\includegraphics[width=6.0in]{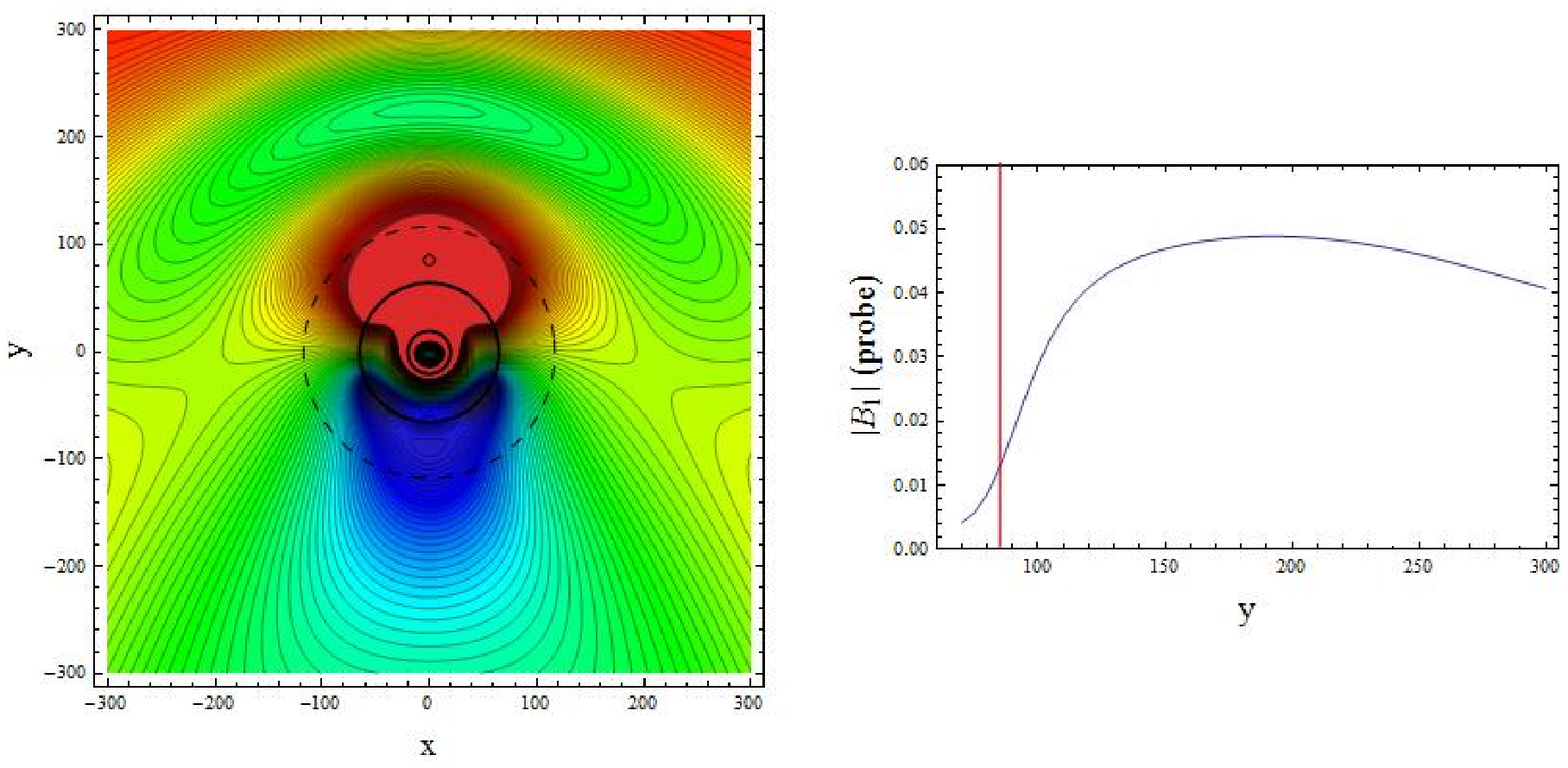}
\caption{\label{fig01x} Left: Contour plot of $|H_\mathrm{z}|$ as a function
of position for a system consisting of a coated cylinder
($r_\mathrm{c} = 20$nm, $r_\mathrm{s} = 65$nm, $\varepsilon_\mathrm{c} = 1$, $\varepsilon_\mathrm{s} = -1 + 0.1\,\mathrm{i}$,
$\mu_\mathrm{c} = \mu_\mathrm{s} = 1$)
interacting with a probe cylinder
($a = 5$nm, $\varepsilon = -1 + 0.1\,\mathrm{i}$, $\mu = 1$), and
irradiated by a $H_\mathrm{z}$ polarized
plane wave with wavelength $600$nm coming from above.
The probe cylinder is within the cloaking region bounded
by the dashed circle, at a distance of 85nm from the origin. Right: magnitude of the dipole moment of the probe cylinder as
a function of its position indicated by the red line. The magnetic field varies in the range
$0.63 \le |H_\mathrm{z}| \le 1.79$, whereas the incident plane wave is normalized to
$|H_{\rm z}^{(inc)}| = 1$ }
\end{figure}

\begin{figure}[h]
\includegraphics[width=6.0in]{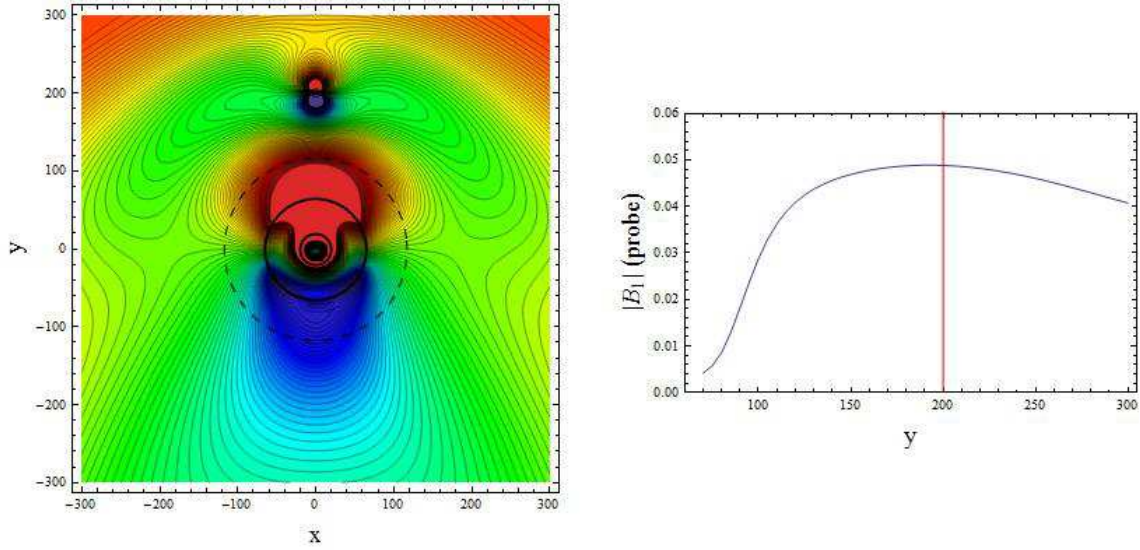}
\caption{\label{fig02x} As for Fig.~\ref{fig01x} with the probe cylinder
outside the cloaking region. Now, the probe is at a distance of 200nm from the origin,
and $0.21 \le |H_\mathrm{z}| \le 1.81$.}
\end{figure}

This physical system is subjected to an incident  plane wave having
$\lambda = 600$nm and with the wave vector in the $xy$-plane
(in-plane incidence) and polarized with
the magnetic field parallel to the cylinder axes ($H_\mathrm{z}$ polarization).

The probe cylinder is polarisable, and  has a dipole moment
proportional to the total electric field at its position. When the
probe cylinder moves along the $y$-axis and enters the cloaking
region, marked by the dashed circle in figures \ref{fig01x} and
\ref{fig02x}, one can see the effect of cloaking, in the sense that
the dipole moment of the probe cylinder tends to zero within the cloaking circle of radius
$r_\#=\surd(r_\mathrm{s}^3/r_\mathrm{c})$ \cite{mnrp,mn} (see figure \ref{fig01x}, right
panel). Consequently, the probe cylinder is successfully cloaked
within $r_\#$ but not of course outside it, compare the left panels
of figures \ref{fig01x} and \ref{fig02x}. Nevertheless, the cloaking
system is not invisible, since the coated cylinder distorts the
incident plane wave. Note that figures \ref{fig01x} and \ref{fig02x} are frames in the
animation available with
this paper, which illustrates the quenching of the dipole moment of
the probe cylinder within the cloaking circle, and its re-emergence
outside it.

The relative permittivity value chosen for the shell in figures \ref{fig01x} and \ref{fig02x}
is comparable to
that of silicon carbide near $\lambda = 10\mu$m \cite{palik},
and has an imaginary part somewhat
lower than that of silver at the wavelength in the ultraviolet where the real part of its
permittivity passes through -1. While we refer to distances in nanometers, in practice
the relevant parameter is the wavelength of the incident radiation divided by a
characteristic length, say the outer radius $r_\mathrm{s}$ of the coated cylinder.
Hence, the results shown in figures \ref{fig01x} and \ref{fig02x} and subsequent figures can be easily
applied to systems rescaled to correspond to materials other than those mentioned.

Note that in the previous figures and simulations we have
presented \cite{mnrp,mn,usoptexp} to illustrate resonant cloaking, the value of
the imaginary part of $\varepsilon_\mathrm{s}$ was chosen to correspond to the mathematical
analysis, rather than to practical materials. We have found that good quenching of
the dipole moment of the probe particle can be achieved even with quite significant
imaginary parts for $\varepsilon_\mathrm{s}$, provided $r_\#$ is sufficiently in excess of $r_\mathrm{s}$,
so that the probe particle can move deep within the cloaking region.

Despite the quite effective cloaking of the probe cylinder shown in the right panels of
figures \ref{fig01x} and \ref{fig02x}, the left panels illustrate strong distortion of the
incident wave in the
vicinity of the cloaking cylinder. Such variations of magnetic field strength would
compromise any attempts to hide the compound system of cloaking cylinder plus probe.

We mention that in all numerical computations we have used the scattering-matrix
method \cite{Felbacq},
which is based on the expansion of the fields in terms of Fourier-Bessel series around
each cylinder.
By using the scattering matrices of each cylinder and the translation properties of
Fourier-Bessel functions, the method leads to the inversion of a linear set of equations.

\section{The quasistatic limit}

For a coated cylinder centered at the origin
of coordinates, we represent the electric and magnetic fields
$E_\mathrm{z}$ and $H_\mathrm{z}$ (denoted here by $V$), by series expansions in terms of
cylindrical harmonics \cite{usarxiv}:
\begin{equation}
V(r, \theta, z, t) = \sum_{\ell=-\infty}^\infty
\left\{
\begin{array}{l}
{\displaystyle A_\ell^\mathrm{c} J_\ell(k_\mathrm{c} r)\, \mathrm{e}^{i \ell \theta}} \\
~\\
{\displaystyle \left[ A_\ell^\mathrm{s} J_\ell(k_\mathrm{s} r) + B_\ell^\mathrm{s} H_\ell^{(1)}(k_\mathrm{s} r) \right]
\mathrm{e}^{i \ell \theta}} \\
~\\
{\displaystyle
\left[ A_\ell^{m} J_\ell(k_\mathrm{m} r) +
 B_\ell^{m} H_\ell^{(1)}(k_\mathrm{m} r) \right]  \mathrm{e}^{i \ell \theta}}
\end{array}
\right\}
\mathrm{e}^{\mathrm{i}\,(\beta\,z - \omega\,t)} \,,
\label{fieldsx}
\end{equation}
where $J_\ell(.)$ and $H_\ell^{(1)}(.)$ represent the Bessel and Hankel functions of the first kind.
The three forms of the series expansions in (\ref{fieldsx}) correspond to the domains
$0 \leq r \leq r_\mathrm{c}$ (inside the core of the coated cylinder),
$r_\mathrm{c} \leq r \leq r_\mathrm{s}$ (inside the shell of the coated cylinder) and
$r \geq r_\mathrm{s}$ (in the matrix), respectively.
Also, the superscripts $c$, $s$ and $m$ label the fields inside the
cylinder core, cylinder shell, and in the matrix, respectively.
Thus, we have the wavenumbers $k_\mathrm{c}^2=\omega^2 \varepsilon_\mathrm{c} \mu_\mathrm{c}$,
$k_\mathrm{s}^2=\omega^2 \varepsilon_\mathrm{s} \mu_\mathrm{s}$, and
$k_\mathrm{m}^2=\omega^2 \varepsilon_\mathrm{m} \mu_\mathrm{m}$.

The function $V$ has to satisfy the boundary conditions, i.e.,
the continuity of the tangential components of the electric
($E_\mathrm{z}$ and $E_\theta$)
and magnetic ($H_\mathrm{z}$ and $H_\theta$) fields across the core
and shell surfaces.
When the coated cylinder is subjected to an incident radiation which is perpendicular
to the axis of the cylinder, we have $\beta=0$,
and the problem can be reduced to solving two independent
problems \cite{panofsky}:
\begin{itemize}
\item
$E_\mathrm{z}$ polarization, when $H_\mathrm{z}=0$ and the transverse
parts of ${\bf H}$ are generated by $\nabla\, E_\mathrm{z}$, and
\item
$H_\mathrm{z}$ polarization, when $E_\mathrm{z}=0$ and
$\nabla\, H_\mathrm{z}$ gives the transverse components of ${\bf E}$.
\end{itemize}

In the present analysis we are interested in the relation between the coefficients in the
matrix, which has the form
\begin{equation}
A_\ell^\mathrm{m}= - M_\ell  B_\ell^\mathrm{m} \, .
\label{a21x}
\end{equation}
The coefficients $A_\ell^\mathrm{m}$ are determined by the sources of the field applied to the structure,
and satisfy the field identity \cite{usoptexp}
\begin{equation}
\sum_{\ell=-\infty}^{\infty}
A_{\ell}^\mathrm{m}\, J_{\ell}(k_\mathrm{m} r) \, \mathrm{e}^{\mathrm{i}\,\ell\,\theta} =
\mathrm{source~field}\,.
\label{eqfi03}
\end{equation}
Hence, we obtain the coefficients $A_{\ell}^\mathrm{m}$ by expanding the source field in terms of
cylindrical harmonics $J_{\ell}(k_\mathrm{m} r) \, \mathrm{e}^{\mathrm{i}\,\ell\,\theta}$.

Here, we also consider that the field applied to the physical structure is a plane wave field.
In cylindrical coordinates, for $H_\mathrm{z}$ polarization
($H_x = H_y = 0$), a magnetic plane wave is described by the formula
\begin{equation}
H_\mathrm{z}^{\mathrm{PW}}(r, \varphi, z) = H_\mathrm{0}\,\mathrm{e}^{\mathrm{i}\,[k_\mathrm{0}\,r\,
\cos{(\varphi - \psi_\mathrm{0}) + k_\mathrm{z}\,z]}}\,,
\label{eqpwf01}
\end{equation}
where $\psi_\mathrm{0}$ is the angle of incidence with respect to the x-axis.
We consider the case of in-plane incidence ($k_\mathrm{z} \equiv$ $\beta = 0$) so that
the exponential in (\ref{eqpwf01}) can be expanded in terms of Bessel functions of the first kind
\begin{equation}
H_\mathrm{z}^{\mathrm{PW}}(r, \varphi) = H_\mathrm{0}
\sum_{n=-\infty}^\infty \mathrm{i}^n\,
J_n(k_\mathrm{0}\,r)\,\mathrm{e}^{\mathrm{i}\,n\,(\varphi - \psi_\mathrm{0})}\,.
\label{eqpwf02}
\end{equation}

Consequently,
for a coated cylinder subjected to a plane wave incoming field, perpendicular to the cylinder axis,
we have the coefficients
\begin{equation}
A_n^\mathrm{m} = H_\mathrm{0}\,\mathrm{i}^n\,\mathrm{e}^{-\mathrm{i}\,n\,\psi_\mathrm{0}}\,,
\label{eqpwf03}
\end{equation}
where
\begin{equation}
\psi_\mathrm{0} = \left\{
\begin{array}{rl}
\pi, & \mathrm{ if~the~radiation~comes~from~} x = +\infty,\\
& \\
0, & \mathrm{ if~the~radiation~comes~from~} x = -\infty.
\end{array} \right.
\label{eqpwf04}
\end{equation}

In the case of $E_\mathrm{z}$ polarization we obtain an equation identical to (\ref{eqpwf03})
for the coefficients of the electric field.

\subsection{$H_\mathrm{z}$ Polarization}
\label{tepr}

We concentrate now on a coated cylinder, centered about the origin of coordinates,
made from non-magnetic materials for which
$\mu_\mathrm{m}=\mu_\mathrm{s}=\mu_\mathrm{c}=\mu_0$, so that
$\varepsilon_\mathrm{m}=n_\mathrm{m}^2 \varepsilon_0$,
$\varepsilon_\mathrm{s}=n_\mathrm{s}^2 \varepsilon_0$ and
$\varepsilon_\mathrm{c}=n_\mathrm{c}^2 \varepsilon_0$, where $\varepsilon_0$
is the dielectric constant of free space, and $n_\mathrm{i} $
(i = m, s, c) represent the refractive indexes of the matrix, shell
and core, respectively.
The boundary conditions coefficients $M_\ell$ from (\ref{a21x}), are derived by eliminating $A_\ell^\mathrm{c}$
in the equations \cite{usarxiv}
\begin{eqnarray}
\fl
\left[ \begin{array}{c} A_\ell^\mathrm{m} \\ B_\ell^\mathrm{m} \end{array} \right]
& = &
\left[ \begin{array}{rr}
J_\ell(k_\mathrm{m} r_\mathrm{s}) & H_\ell(k_\mathrm{m} r_\mathrm{s}) \\
Z_\mathrm{m} J_\ell'(k_\mathrm{m} r_\mathrm{s}) & Z_\mathrm{m} H_\ell'(k_\mathrm{m} r_\mathrm{s}) \end{array}
\right]^{-1}
\left[ \begin{array}{rr}
J_\ell(k_\mathrm{s} r_\mathrm{s}) & H_\ell(k_\mathrm{s} r_\mathrm{s}) \\
Z_\mathrm{s} J_\ell'(k_\mathrm{s} r_\mathrm{s}) & Z_\mathrm{s} H_\ell'(k_\mathrm{s} r_\mathrm{s}) \end{array}
\right]
\label{TEbcd} \\ & \times &
\left[ \begin{array}{rr}
J_\ell(k_\mathrm{s} r_\mathrm{c}) & H_\ell(k_\mathrm{s} r_\mathrm{c}) \\
Z_\mathrm{s} J_\ell'(k_\mathrm{s} r_\mathrm{c}) & Z_\mathrm{s} H_\ell'(k_\mathrm{s} r_\mathrm{c}) \end{array}
\right]^{-1}
\left[ \begin{array}{rr}
J_\ell(k_\mathrm{c} r_\mathrm{c}) & H_\ell(k_\mathrm{c} r_\mathrm{c}) \\
Z_\mathrm{c} J_\ell'(k_\mathrm{c} r_\mathrm{c}) & Z_\mathrm{c} H_\ell'(k_\mathrm{c} r_\mathrm{c}) \end{array}
\right]
\left[ \begin{array}{c} A_\ell^\mathrm{c} \\ 0 \end{array} \right]
\, ,
\nonumber
\end{eqnarray}
where $J_\ell(.)$ and $H_\ell(.) \equiv H_\ell^{(1)}(.)$ are Bessel and Hankel
functions of the first kind, the prime indicates the derivative of the
corresponding function, and $Z_\mathrm{i} =\sqrt{\mu_\mathrm{i} /\varepsilon_\mathrm{i} }$
(i = m, s, c) represent the impedances of the matrix, shell
and core, respectively.
For $E_\mathrm{z}$ polarization we obtain the relation between $A_\ell^\mathrm{m}$
and $B_\ell^\mathrm{m}$ by changing $Z_\mathrm{i} \rightarrow1/Z_\mathrm{i} $ in (\ref{TEbcd}).

In the quasistatic limit ($k_\mathrm{m}\rightarrow 0$), we approximate the Bessel functions by the
first term in their series expansion, i.e.
\begin{eqnarray*}
J_0(z) &\approx & 1 - \left(\frac{z}{2}\right)^2 , \\
&& \\
J_{\mathrm{n}}(z) & \approx & \left\{ \begin{array}{lcl}
{\displaystyle \frac{1}{n!}\, \left( \frac{z}{2} \right)^n } & {\rm ~for~} & n \geq 0 , \\
&& \\
{\displaystyle (-1)^n\,\frac{1}{(-n)!}\, \left( \frac{z}{2} \right)^{-n} } & {\rm ~for~} & n < 0 ,
\end{array} \right.
\end{eqnarray*}
\begin{eqnarray*}
H_0(z) &\approx & J_0(z) + \mathrm{i}\,\frac{2}{\pi}
\left[\gamma^E + \log{\left(\frac{z}{2}\right)} \right], \\
&& \\
H_{\mathrm{n}}(z) & \approx & J_\mathrm{n}(z) + {\rm i}\,\left\{ \begin{array}{lcl}
{\displaystyle - \frac{1}{\pi}\, \left( \frac{2}{z} \right)^n (n -1)!} & {\rm ~for~} & n \geq 0 ,\\
&& \\
{\displaystyle (-1)^{n+1}\,\frac{1}{\pi}\, \left( \frac{2}{z} \right)^{-n}  (-n -1)!} &
{\rm ~for~} & n < 0 ,
\end{array} \right.
\end{eqnarray*}
where $\gamma^E$ is the Euler-Mascheroni constant \cite{AS}.
Then, we substitute these expressions in
(\ref{TEbcd}) and write $M_\ell$ as a fraction. Using the limit $k_\mathrm{m} \rightarrow 0$ we determine
the coefficient of $k_\mathrm{m}^0$ in the numerator and the coefficient of $k_\mathrm{m}^{2\ell}$ in the denominator.
Thus, the quasistatic limit of $M_\ell$ for $\ell \geq 0$ is
\begin{equation}
M_\ell = M_{-\ell} \approx
- \frac{\mathrm{i}}{\pi} \left( \frac{2}{k_\mathrm{m} r_\mathrm{s}} \right)^{2 \ell}
\ell! (\ell-1)! \, \gamma_\ell \, ,
\label{eq6ad}
\end{equation}
where
\begin{equation}
\gamma_\ell =
\frac{r_\mathrm{c}^{2\ell} (\varepsilon_\mathrm{s} - \varepsilon_\mathrm{c})
(\varepsilon_\mathrm{m} - \varepsilon_\mathrm{s}) + r_\mathrm{s}^{2\ell}
(\varepsilon_\mathrm{s} + \varepsilon_\mathrm{c})
(\varepsilon_\mathrm{m} + \varepsilon_\mathrm{s})}
{r_\mathrm{c}^{2\ell} (\varepsilon_\mathrm{s} - \varepsilon_\mathrm{c})
(\varepsilon_\mathrm{m} + \varepsilon_\mathrm{s}) + r_\mathrm{s}^{2\ell}
(\varepsilon_\mathrm{s} + \varepsilon_\mathrm{c})
(\varepsilon_\mathrm{m} - \varepsilon_\mathrm{s})} \, .
\label{eq6a}
\end{equation}
For $\ell=0$, we obtain a completely different form
\begin{equation}
M_0 \approx - \frac{\mathrm{i}}{\pi} \left( \frac{2}{k_\mathrm{m} r_\mathrm{s}} \right)^4
\frac{\varepsilon_\mathrm{m}}{(\varepsilon_\mathrm{s}-\varepsilon_\mathrm{c})
(r_\mathrm{c}/r_\mathrm{s})^4 + (\varepsilon_\mathrm{m}-\varepsilon_\mathrm{s})} \, .
\label{tem0}
\end{equation}
Note that in all these calculations we made no assumption about the nature (real or complex)
of permittivities or refractive indices.

To relate the long wavelength limit of the dynamic problem with
the corresponding problem in electrostatics we apply the same
method as in Ref.~\cite{jewatem}. Thus, the boundary conditions for our problem
correspond to an electrostatic problem in which the inverse of the dielectric constants
($\varepsilon_\mathrm{c}\rightarrow 1/\varepsilon_\mathrm{c}$,
$\varepsilon_\mathrm{s}\rightarrow 1/\varepsilon_\mathrm{s}$ and
$\varepsilon_\mathrm{m}\rightarrow 1/\varepsilon_\mathrm{m}$) have to be considered.
This will also change
$\gamma_\ell\rightarrow -\gamma_\ell$. Now, the boundary conditions
(\ref{a21x}) for $\ell\neq0$ can be written in the form
\begin{equation}
A_\ell^\mathrm{m} \approx - \frac{\mathrm{i}}{\pi} \left( \frac{2}{k_\mathrm{m}} \right)^{2 \ell}
\ell! (\ell-1)! \, \gamma_\ell \,
\frac{B_\ell^\mathrm{m}}{r_\mathrm{s}^{2\ell}} .
\label{tepr1}
\end{equation}
Note that, here, we separated the product $k_\mathrm{m} r_\mathrm{s}$, which is dimensionless, so that
$k_\mathrm{m}$ and $r_\mathrm{s}$ are considered as multiplied, respectively divided, by a length unit.

In electrostatics, the corresponding relationship between the
coefficients $A_\ell$ and
$B_\ell$ which controls the response of a coated cylinder to an
external field, has the form \cite{cc1,cc2}
\begin{equation}
\widetilde{A}_\ell = \gamma_\ell \, \frac{\widetilde{B}_\ell}{r_\mathrm{s}^{2\ell}} \, .
\label{statics}
\end{equation}
Now, by comparing (\ref{tepr1}) with (\ref{statics}) we may
infer the relation between static and dynamic multipole coefficients
\begin{equation}
\widetilde{B}_\ell \approx - \frac{\mathrm{i}}{\pi} \left( \frac{2}{k_\mathrm{m}} \right)^{2 \ell}
\ell! (\ell-1)! \, B_\ell^\mathrm{m}
\approx \frac{H_\ell(k_\mathrm{m})}{J_\ell(k_\mathrm{m})}\, B_\ell^\mathrm{m} \,,
\quad \mathrm{~for~} \ell\neq0.
\label{tepr2}
\end{equation}
Note that $k_\mathrm{m}$ is dimensionless according to the
note after (\ref{tepr1}).


In electrostatics, the partial resonances of a three--phase
composite consisting of coated cylinders are defined by the
equations \cite{cc1,cc2}
\begin{eqnarray}
\varepsilon_\mathrm{c} + \varepsilon_\mathrm{s} & = & 0 \qquad ({\rm core-shell~
 resonance}),
\label{cspr} \\
\varepsilon_\mathrm{s} + \varepsilon_\mathrm{m} & = & 0 \qquad ({\rm shell-matrix~
 resonance}),
\label{smpr}
\end{eqnarray}
when (\ref{statics}) becomes
\begin{equation}
\widetilde{A}_\ell = \frac{\varepsilon_\mathrm{m} + \varepsilon_\mathrm{c}}{\varepsilon_\mathrm{m} -
\varepsilon_\mathrm{c}} \, \frac{\widetilde{B}_\ell}{r_\mathrm{s}^{2\ell}} \, ,
\label{cspres}
\end{equation}
or
\begin{equation}
\widetilde{A}_\ell = \frac{\varepsilon_\mathrm{m} + \varepsilon_\mathrm{c}}{\varepsilon_\mathrm{m} -
\varepsilon_\mathrm{c}} \, \frac{\widetilde{B}_\ell}{(r_\mathrm{s}^2/r_\mathrm{c})^{2\ell}} \, ,
\label{smpres}
\end{equation}
respectively.
In the first case, the field inside the coated cylinder is exactly
the same as would be found within a solid cylinder of radius $r_\mathrm{s}$
and dielectric constant $\varepsilon_\mathrm{c}$, while the potential
outside the coated cylinder, in the matrix, is precisely the same
as that outside the solid cylinder \cite{mnrp,cc1}.
The second case corresponds to a
solid cylinder of radius $r_\mathrm{s}^2/r_\mathrm{c} > r_\mathrm{s}$ (the geometrical image of
the core boundary with respect to the shell outer boundary), and
dielectric constant $\varepsilon_\mathrm{c}$. Now, the field external to the coated
cylinder {\em and beyond the radius} $r_* = r_\mathrm{s}^2/r_\mathrm{c}$ is the same as that
external to the solid cylinder \cite{mnrp,cc1}.

Since it is the relationship between the coefficients $A_\ell^\mathrm{m}$ and
$B_\ell^\mathrm{m}$ which controls the response of a coated cylinder to an
external field,
equations (\ref{statics}) and (\ref{tepr1}) show that this
response is determined by $\gamma_\ell$ in electrostatics as well
as in the long wavelengths limit of electrodynamics.
The limiting process is smooth and
therefore, we expect a resonant behavior accompanied by cloaking effects,
even for nonzero frequencies, when one of the
conditions (\ref{cspr}) or (\ref{smpr}) is satisfied.

\subsection{$E_\mathrm{z}$ Polarization}
\label{tmpr}

Now, in the long wavelength limit,
the boundary conditions coefficients $M_\ell$ from (\ref{a21x})
take the form
\begin{eqnarray}
M_\ell  =  M_{-\ell}
 \approx - \frac{\mathrm{i}}{\pi} \left( \frac{2}{k_\mathrm{m} r_\mathrm{s}} \right)^{2 \ell + 2}
\frac{2 \, \ell! (\ell+1)! \, \varepsilon_\mathrm{m}}{(\varepsilon_\mathrm{s} - \varepsilon_\mathrm{c})
(r_\mathrm{c}/r_\mathrm{s})^{2\ell+2} + (\varepsilon_\mathrm{m} - \varepsilon_\mathrm{s})}  ,
\label{tmpr1}
\end{eqnarray}
for $\ell\neq0$, and
\begin{equation}
M_0 \approx - \frac{\mathrm{i}}{\pi} \left( \frac{2}{k_\mathrm{m} r_\mathrm{s}} \right)^{2}
\frac{\varepsilon_\mathrm{m}}{(\varepsilon_\mathrm{s} - \varepsilon_\mathrm{c})
(r_\mathrm{c}/r_\mathrm{s})^{2} + (\varepsilon_\mathrm{m} - \varepsilon_\mathrm{s})} \, .
\label{tmpr2}
\end{equation}

Note that there are no terms of the form $\varepsilon_\mathrm{c} + \varepsilon_\mathrm{s}$ or
$\varepsilon_\mathrm{s} + \varepsilon_\mathrm{m}$ in (\ref{tmpr1}) or (\ref{tmpr2})
to indicate a core--shell or shell--matrix partial resonance.
Again, the limiting process is smooth and,
consequently, we do not expect a resonant behaviour of the
coated cylinder, for any frequency, in the case of $E_\mathrm{z}$ polarization.
Consequently, when the coated cylinder is irradiated with a field of a general polarization,
that is a mixture of $H_\mathrm{z}$ and $E_\mathrm{z}$ polarizations, or in the case of
conical incidence, the cloaking by resonance will never be perfect, or even may be
completely ruined, due to the contribution of the $E_\mathrm{z}$ polarized component.

Resonances similar to those in Sec. \ref{TEbcd} can occur in the case of a coated
cylinder made from magnetic metamaterials
with permitivitty $\varepsilon_0$ and permeabilities $\mu_\mathrm{c}$, $\mu_\mathrm{s}$, and $\mu_\mathrm{m}$.
Now, the boundary conditions coefficients $M_\ell$ from (\ref{a21x})
take the forms (\ref{eq6ad}) and (\ref{tem0}) with $\varepsilon_\mathrm{i} $ replaced by $\mu_\mathrm{i} $
\cite{usarxiv,ojfm}. Hence, the magnetic partial resonances of coated cylinders are defined by the
equations
\begin{eqnarray}
\mu_\mathrm{c} + \mu_\mathrm{s} & = & 0 \qquad ({\rm core-shell~resonance}),
\label{csprm} \\
\mu_\mathrm{s} + \mu_\mathrm{m} & = & 0 \qquad ({\rm shell-matrix~resonance}).
\label{smprm}
\end{eqnarray}

\section{Attempts to minimize the Ostrich Effect}

As a measure of effectiveness of cloaking we choose the total scattering cross section.
For two-dimensional problems, the total scattering cross section is defined as the
ratio of the total power scattered by an object, to the incident power per unit
length \cite{panofsky,vanbladel}
\begin{equation}
\sigma_\mathrm{t} = \frac{4}{k_\mathrm{m}} \sum_{\ell = -\infty}^\infty \left| B_\ell^\mathrm{m} \right|^2
= \frac{4}{k_\mathrm{m}}\,\left| B_0^\mathrm{m} \right|^2
+ \frac{8}{k_\mathrm{m}} \sum_{\ell = 1}^\infty \left| B_\ell^\mathrm{m} \right|^2.
\label{sigmat}
\end{equation}
Here, the $B_\ell^\mathrm{m}$ coefficients have the exact form given by (\ref{TEbcd}), that is
\begin{equation}
B_\ell^\mathrm{m} = \frac{P}{Q}\,A_\ell^\mathrm{m} ,
\label{BlmTE}
\end{equation}
where
\begin{eqnarray*}
P & = & \left\{ \left[ Z_\mathrm{m}\, J_\ell'(k_\mathrm{m} r_\mathrm{s})\, J_\ell(k_\mathrm{s} r_\mathrm{s}) - Z_\mathrm{s}\, J_\ell(k_\mathrm{m} r_\mathrm{s})\, J_\ell'(k_\mathrm{s} r_\mathrm{s}) \right] \right. \\
& & \times \left. \left[ Z_\mathrm{c}\, H_\ell(k_\mathrm{s} r_\mathrm{c})\, J_\ell'(k_\mathrm{c} r_\mathrm{c}) - Z_\mathrm{s}\, H_\ell'(k_\mathrm{s} r_\mathrm{c})\, J_\ell(k_\mathrm{c} r_\mathrm{c}) \right] \right\} \\
& - & \left\{ \left[ Z_\mathrm{s}\, H_\ell'(k_\mathrm{s} r_\mathrm{s})\, J_\ell(k_\mathrm{m} r_\mathrm{s}) - Z_\mathrm{m}\, H_\ell(k_\mathrm{s} r_\mathrm{s})\, J_\ell'(k_\mathrm{m} r_\mathrm{s}) \right] \right. \\
& & \times \left. \left[ Z_\mathrm{s}\, J_\ell'(k_\mathrm{s} r_\mathrm{c})\, J_\ell(k_\mathrm{c} r_\mathrm{c}) - Z_\mathrm{c}\, J_\ell(k_\mathrm{s} r_\mathrm{c})\, J_\ell'(k_\mathrm{c} r_\mathrm{c}) \right] \right\}, \\
&& \\
Q & = & \left\{ \left[ Z_\mathrm{m}\, H_\ell'(k_\mathrm{m} r_\mathrm{s})\, J_\ell(k_\mathrm{s} r_\mathrm{s}) - Z_\mathrm{s}\, H_\ell(k_\mathrm{m} r_\mathrm{s})\, J_\ell'(k_\mathrm{s} r_\mathrm{s}) \right] \right. \\
& & \times \left. \left[ Z_\mathrm{s}\, H_\ell'(k_\mathrm{s} r_\mathrm{c})\, J_\ell(k_\mathrm{c} r_\mathrm{c}) - Z_\mathrm{c}\, H_\ell(k_\mathrm{s} r_\mathrm{c})\, J_\ell'(k_\mathrm{c} r_\mathrm{c}) \right] \right\} \\
& - & \left\{ \left[ Z_\mathrm{m}\, H_\ell'(k_\mathrm{m} r_\mathrm{s})\, H_\ell(k_\mathrm{s} r_\mathrm{s}) - Z_\mathrm{s}\, H_\ell(k_\mathrm{m} r_\mathrm{s})\, H_\ell'(k_\mathrm{s} r_\mathrm{s}) \right] \right. \\
& & \times \left. \left[ Z_\mathrm{s}\, J_\ell'(k_\mathrm{s} r_\mathrm{c})\, J_\ell(k_\mathrm{c} r_\mathrm{c}) - Z_\mathrm{c}\, J_\ell(k_\mathrm{s} r_\mathrm{c})\, J_\ell'(k_\mathrm{c} r_\mathrm{c}) \right] \right\} ,
\end{eqnarray*}
and with $A_\ell^\mathrm{m}$ from (\ref{eqpwf03}) for $E_0 = 1$ and $\psi = 0$.

\begin{figure}[h]
\centering
\includegraphics[width=3.5in]{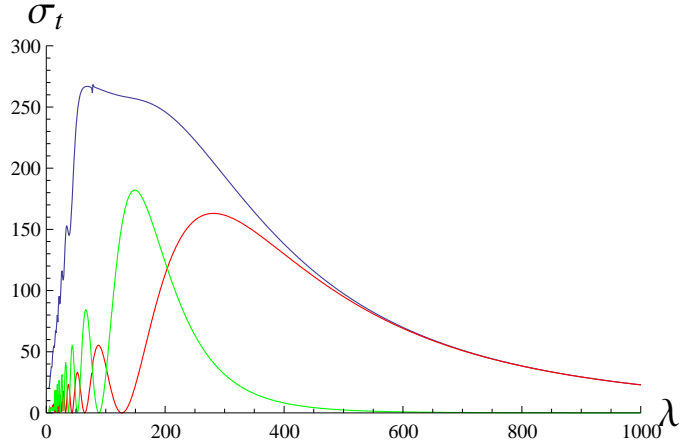}
\caption{\label{fig03x} Total cross section $\sigma_\mathrm{t}^{(6)}$  (blue curve),
$\sigma_\mathrm{t}^{(0)}$ (red curve) from (\ref{numsim}), and
$\sigma_\mathrm{t}^{(1)}-\sigma_\mathrm{t}^{(0)}$ (green curve)
as functions of the wavelength of the incident plane wave,
for a coated cylinder ($r_\mathrm{c} = 20$nm, $r_\mathrm{s} = 65$nm) at resonance
($\varepsilon_\mathrm{c} = \varepsilon_\mathrm{m} = 1$, $\varepsilon_\mathrm{s} = -1$,
$\mu_\mathrm{c} = \mu_\mathrm{s} = \mu_\mathrm{m} = 1$  ).}
\end{figure}

In the case of the resonance $\varepsilon_\mathrm{c} = \varepsilon_\mathrm{m} = 1$ and $\varepsilon_\mathrm{s} = -1$,
numerical simulations show that by using the form (\ref{BlmTE}) and the series
(\ref{sigmat}) truncated to $N_{\rm trunc} = 6$, we have
\begin{equation}
\sigma_\mathrm{t}^{(N_{\rm trunc})} = \frac{4}{k_\mathrm{m}} \sum_{\ell = -N_{\rm trunc}}^{N_{\rm trunc}} \left| B_\ell^\mathrm{m} \right|^2
\approx \frac{4}{k_\mathrm{m}}\,\left| B_0^\mathrm{m} \right|^2 = \sigma_\mathrm{t}^{(0)},
\label{numsim}
\end{equation}
starting at about $\lambda = 10\, r_\mathrm{s}$ (see figure \ref{fig03x}). From the same wavelength up,
the contribution of the dipole terms, given by $\sigma_\mathrm{t}^{(1)} - \sigma_\mathrm{t}^{(0)}$, becomes very small.

If we use the expression of $B_\ell^\mathrm{m}$ in the quasistatic limit (see Sec.~\ref{tepr}),
with $A_\ell^\mathrm{m}$ from (\ref{eqpwf03}), $E_0 = 1$, and $\psi = 0$, we obtain
\begin{equation}
B_\ell^\mathrm{m} \approx \left\{ \begin{array}{lcl}
{\displaystyle \mathrm{i}\,\pi^5\,\frac{(\varepsilon_\mathrm{s}-\varepsilon_\mathrm{c})
(r_\mathrm{c}/r_\mathrm{s})^4 + (\varepsilon_\mathrm{m}-\varepsilon_\mathrm{s})}{\varepsilon_\mathrm{m}} \,
\left( \frac{r_\mathrm{s}}{\lambda} \right)^4 ,} & \mathrm{~for~} & \ell = 0 ,\\
&& \\
{\displaystyle \mathrm{i}^{l+1}\, \pi^{2 \ell +1}\,\frac{1}{\ell!\,(\ell -1)!\,\gamma_\ell}\,
\left( \frac{r_\mathrm{s}}{\lambda} \right)^{2 \ell} ,} & \mathrm{~for~} & \ell \geq 1 .
\end{array} \right.
\label{sig02}
\end{equation}
In the case of core--shell--matrix resonance, that is $\varepsilon_\mathrm{c} +\varepsilon_\mathrm{s} =0$ and
$\varepsilon_\mathrm{s} +\varepsilon_\mathrm{m} =0$, the coefficient $\gamma_\ell$ defined in (\ref{eq6a})
tends to infinity so that, for $\ell \geq 1$, $B_\ell^\mathrm{m} \rightarrow 0$.
Such a situation arises when
\begin{equation}
\varepsilon_\mathrm{c} = -\varepsilon_\mathrm{s} = \varepsilon_\mathrm{m}
\label{sig03}
\end{equation}
(as in the case of $\varepsilon_\mathrm{c} = \varepsilon_\mathrm{m} = 1$ and
$\varepsilon_\mathrm{s} = -1$), and the total scattering cross section is
determined only by the zeroth-order multipole
\begin{equation}
\sigma_\mathrm{t}^{\mathrm{QS}} \approx \frac{4}{k_\mathrm{m}}\,\left| B_0^\mathrm{m} \right|^2
\propto \left( \frac{r_\mathrm{s}}{\lambda} \right)^7 ,
\label{sig04}
\end{equation}
which tends rapidly to zero as the wavelength increases. Actually,
$B_0^\mathrm{m}$ from (\ref{BlmTE}) tends very slowly to the form
(\ref{sig02}). This last form has been obtained by taking the first
term in the series of all Bessel functions, except $J_0(z)$ and
$H_0(z)$). For complicated expressions like (\ref{BlmTE}) the series
expansions require more terms for accuracy, as they contain products
of four Bessel functions.

The main result is that for long wavelengths $\sigma_t$ is determined by the $B_0^\mathrm{m}$, only.
The dominance of zeroth-order multipole is also present in the
case of coordinate transformation method \cite{pendry}. This case has been analyzed by
Yan \etal \cite{qiu}.

Here, we have considered that $\varepsilon_\mathrm{s} = -1$ is real, which is unphysical.
Physical materials with negative permittivity (usually metals) are lossy so that we have to consider
a coated cylinder with the shell material having a complex permittivity
$\varepsilon_\mathrm{s} = -1 + {\rm i}\,\delta$, where $\delta > 0$ determines the loss in the shell.
A detailed analysis of this case shows that, in the limit of long wavelengths,
the total cross section $\sigma_t$ is now dominated by the dipole coefficients $B_{\pm 1}^\mathrm{m}$.

\begin{figure}[h]
\centering
\includegraphics[width=3.5in]{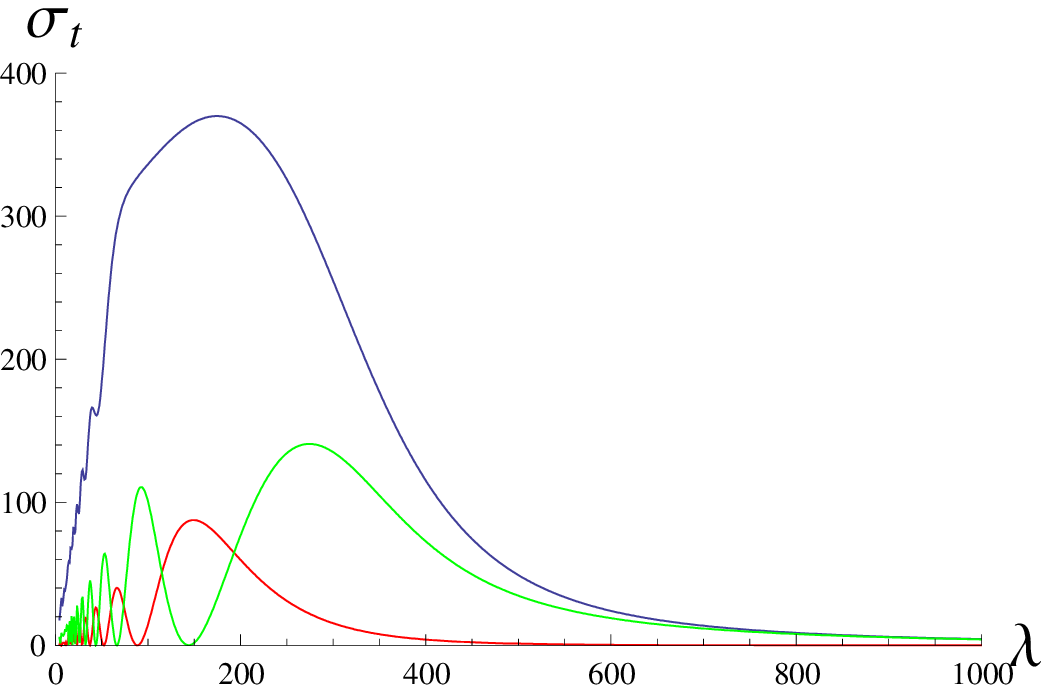}
\caption{\label{fig04x} Total cross section $\sigma_\mathrm{t}^{(6)}$  (blue curve),
$\sigma_\mathrm{t}^{(0)}$ (red curve) from (\ref{numsim}), and
$\sigma_\mathrm{t}^{(1)}-\sigma_\mathrm{t}^{(0)}$ (green curve)
as functions of the wavelength of the incident plane wave,
for a coated cylinder ($r_\mathrm{c} = 20$nm, $r_\mathrm{s} = 65$nm) at resonance
($\varepsilon_\mathrm{c} = \varepsilon_\mathrm{m} = 1$, $\varepsilon_\mathrm{s} = -1+0.1\,\mathrm{i}$,
$\mu_\mathrm{c} = \mu_\mathrm{s} = \mu_\mathrm{m} = 1$  ).}
\end{figure}

We start with the analytic form of $B_\ell^\mathrm{m}$ coefficients (\ref{BlmTE}) and set the factor $A_\ell^\mathrm{m} = 1$.
In fact, $A_\ell^\mathrm{m}$ defined in (\ref{eqpwf03}) has the modulus $|A_\ell^\mathrm{m}| = E_0^2$ and we consider$E_0 = 1$.
Now, for a finite $\delta$, from the exact form (\ref{BlmTE}) we obtain the following series expansion
\begin{equation}
B_0^\mathrm{m} = a_4(\delta)\,k_0^4 + a_6(\delta)\,k_0^6 +
\mathcal{O}(k_0^8)\,, \label{sig05}
\end{equation}
where
\begin{eqnarray}
\fl
a_4(\delta) & = & \mathrm{i} \frac{\pi}{16} ( r_\mathrm{c}^4 - r_\mathrm{s}^4 ) +
\frac{\pi}{32} ( r_\mathrm{c}^4 - r_\mathrm{s}^4 ) \delta\,,
\label{sig061} \\
\fl
a_6(\delta) & = & \mathrm{i} \frac{\pi}{32} ( r_\mathrm{c}^6 - 2 r_\mathrm{c}^4 r_\mathrm{s}^2 + r_\mathrm{s}^6 )
+
\frac{\pi}{192} ( 7 r_\mathrm{c}^6 - 12 r_\mathrm{c}^4 r_\mathrm{s}^2 + 5 r_\mathrm{s}^6 ) \delta
 -
\mathrm{i} \frac{\pi}{192} ( 2 r_\mathrm{c}^6 - 3 r_\mathrm{c}^4 r_\mathrm{s}^2 + r_\mathrm{s}^6 ) \delta^2.
\nonumber \\
\fl &&
\label{sig062}
\end{eqnarray}
It is easy to check that the first term in the limit
\begin{equation}
\lim_{\delta \rightarrow 0} B_0^\mathrm{m} = \mathrm{i}\,\frac{\pi}{16}
\left( r_\mathrm{c}^4 - r_\mathrm{s}^4 \right) k_0^4 + \mathrm{i}\,\frac{\pi}{32}
\left( r_\mathrm{c}^6 - 2 r_\mathrm{c}^4 r_\mathrm{s}^2 + r_\mathrm{s}^6 \right) k_0^6 +
\mathcal{O}(k_0^8)\,, \label{sig07}
\end{equation}
is of the form (\ref{sig02}) for $\ell = 0$, if we set
$\varepsilon_\mathrm{c} = \varepsilon_\mathrm{m} = 1$ and
$\varepsilon_\mathrm{s} = -1$ in (\ref{sig02}).
Consequently, the behaviour of $B_0^\mathrm{m}$ as function of $k_0$ and $\delta$, in the domain of
long wavelengths, can be summarized as
\begin{equation}
B_0^\mathrm{m} \simeq\left\{
\begin{array}{lcl}
a_4(\delta)\,k_0^4 & \mathrm{~~if~~} & \delta \neq 0\,, \\
&& \\
a_4(0)\,k_0^4 & \mathrm{~~if~~} & \delta = 0\,.\end{array}
\right.
\label{sig08}
\end{equation}

Now, we analyze the dipole term $B_1^\mathrm{m}$. Firstly, from (\ref{BlmTE}) we obtain
\begin{equation}
B_1^\mathrm{m} = b_2(\delta)\,k_0^2 + b_4(\delta)\,k_0^4 +
\mathcal{O}(k_0^6)\,, \label{sig09}
\end{equation}
where
\begin{eqnarray}
\fl
b_2(\delta)  = \frac{\pi}{4}\, \frac{r_\mathrm{s}^2 (r_\mathrm{s}^2 - r_\mathrm{c}^2) (2 -
\mathrm{i}\,\delta)}{- r_\mathrm{s}^2 \delta^2 + r_\mathrm{c}^2 (2 +
\mathrm{i}\,\delta)}\,\delta = \frac{\pi\,rs^2 (r_\mathrm{c}^2 -
r_\mathrm{s}^2)}{8\,r_\mathrm{c}^2}\,\delta +
\mathrm{i}\,\frac{\pi\,r_\mathrm{s}^2}{16\,r_\mathrm{c}^2}\,(r_\mathrm{c}^2 - r_\mathrm{s}^2) \delta^2 +
\mathcal{O}(\delta^3)\,,
\label{sig10} \\
\fl
b_4(\delta) = - \mathrm{i}\,\frac{\pi}{8}\,r_\mathrm{s}^4\,\log{\left(
\frac{r_\mathrm{c}}{r_\mathrm{s}} \right)} - \frac{\pi}{8}\,r_\mathrm{s}^4\,\log{\left(
\frac{r_\mathrm{c}}{r_\mathrm{s}} \right)}\,\delta + \mathcal{O}(\delta^2)\,.
\label{sig11}
\end{eqnarray}
We also have
\begin{equation}
\lim_{\delta \rightarrow 0} B_1^\mathrm{m} \simeq -
\mathrm{i}\,\frac{\pi}{8}\,r_\mathrm{s}^4\,\log{\left( \frac{r_\mathrm{c}}{r_\mathrm{s}}
\right)} k_0^4\,, \label{sig12}
\end{equation}
Finally, the behaviour of $B_1^\mathrm{m}$ as function of $k_0$ and $\delta$, in the domain of
long wavelengths, can be summarized as
\begin{equation}
B_1^\mathrm{m} \simeq \left\{
\begin{array}{lcl}
b_2(\delta)\,k_0^2 & \mathrm{~~if~~} & \delta \neq 0\,, \\
&& \\
b_4(0)\,k_0^4 & \mathrm{~~if~~} & \delta = 0\,.\end{array}
\right.
\label{sig13}
\end{equation}

It follows that, in the long wavelength limit, the scattering cross
section is dominated by the monopole term $\sigma_\mathrm{t}^{(0)}=4\,|B_0^\mathrm{m}|^2/k_\mathrm{m}$ if $\delta =
0$, and by the dipole term
$\sigma_\mathrm{t}^{(1)}-\sigma_\mathrm{t}^{(0)}=8\,|B_1^\mathrm{m}|^2/k_\mathrm{m}$ when $\delta \neq 0$. This
last result agrees with that obtained by Alu and Engheta
\cite{engheta} who have also shown that the
dipole term dominates the scattering cross section, for lossy materials.

\begin{figure}[h]
\centering
\includegraphics[width=3.0in]{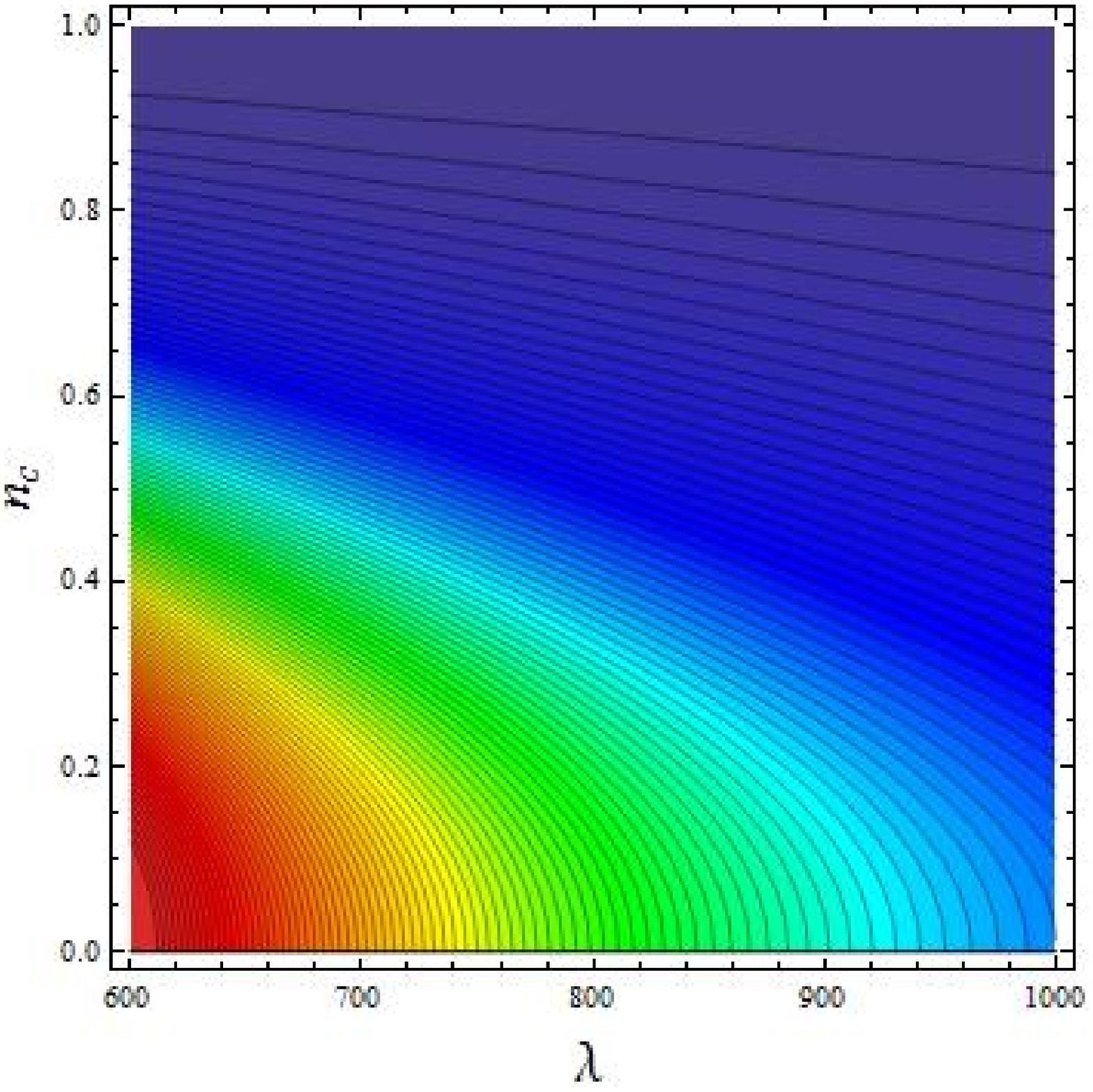}~~~
\includegraphics[width=3.0in]{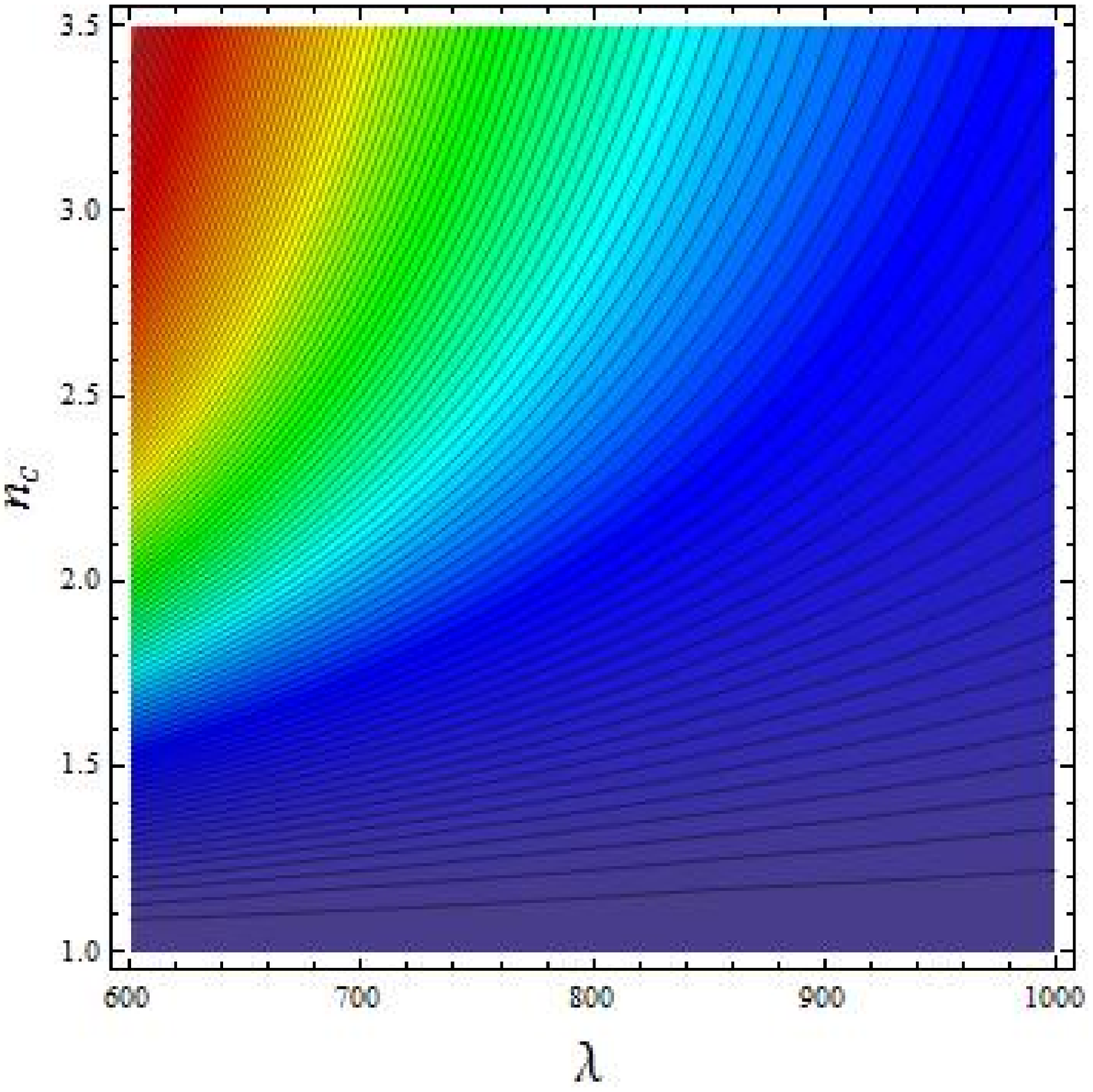}
\caption{\label{fig056x} Scattering cross section $\sigma_{\mathrm{t}}^{(6)}$, from
equation (\ref{sigmat}), as a function of wavelength ($\lambda$) and
core refractive index ($n_\mathrm{c}$). Left: $n_\mathrm{c} <1$. Right: $n_\mathrm{c}>1$.
The parameters of the physical system are: $r_\mathrm{c} = 20$nm, $r_\mathrm{s} =
65$nm, $\varepsilon_\mathrm{c} = 1$, $\varepsilon_\mathrm{s} = -1 + 0.1\,\mathrm{i}$,
$\mu_\mathrm{c} = \mu_\mathrm{s} = 1$. Red indicates large cross sections (up to 18.0),
while blue indicates smaller values (down to 0.13).}
\end{figure}

\begin{figure}[h]
\centering
\includegraphics[width=4.0in]{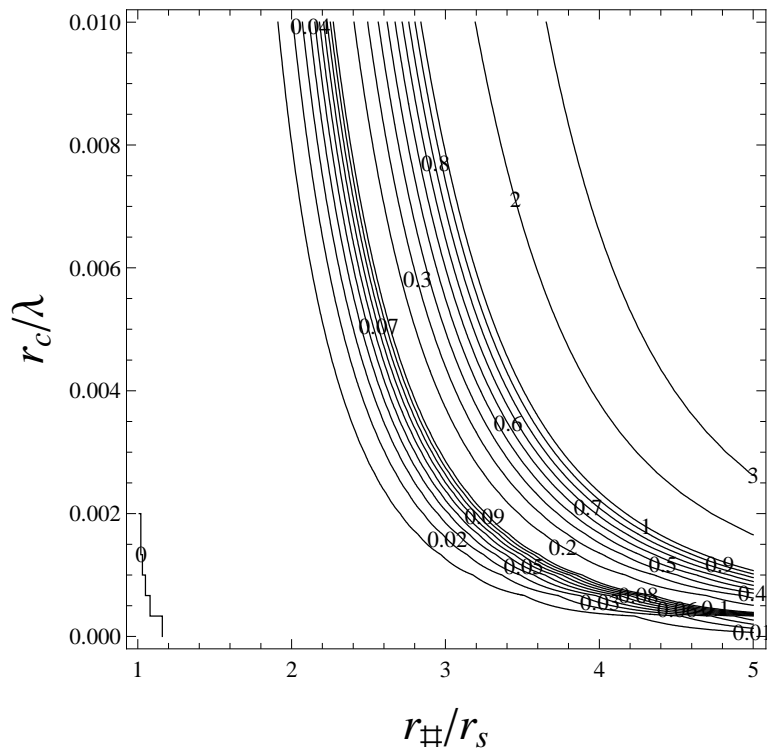}
\caption{\label{fig07x} Normalized scattering cross section
$\sigma_{\mathrm{t}}^{(6)}/2 r_\mathrm{s}$, from equation (\ref{sigmat}), as a function of
normalized core radius ($r_\mathrm{c}/\lambda$) and $r_\#/r_\mathrm{s}=\sqrt{r_\mathrm{s}/r_\mathrm{c}}$.
Here $\varepsilon_\mathrm{c} = 1$, $\varepsilon_\mathrm{s} = -1 + 0.1\,\mathrm{i}$,
$\mu_\mathrm{c} = \mu_\mathrm{s} = 1$.}
\end{figure}

Figure \ref{fig04x} shows the cross section $\sigma_\mathrm{t}$ as a function of
wavelength calculated now for a realistic value $\epsilon_\mathrm{s}=-1+0.1
i$ of the dielectric constant of the shell. In comparison with figure \ref{fig03x},
we see that the dominant contribution to the cross section now
comes from the dipole terms rather than the monopole terms. Despite
this difference, the cross section is well approximated by its
leading term when the wavelength reaches around 10 times the shell
radius. Note that in figure \ref{fig03x} the cross section varies as
approximately $1/\lambda^2$ in the region of $\lambda$ between 600
and 1000; this is far from the quasistatic behaviour of
$1/\lambda^7$ expected from equation (\ref{sig08}), showing that for
this ideal case the cross section is not well represented by
quasistatics even at $\lambda=1000$. By contrast, for figure \ref{fig04x} the
cross section goes as $1/\lambda^3$, in line with the quasistatic
estimate (see equation (\ref{sig13}), and also Panicky and Phillips \cite{panofsky}).

\begin{figure}[h]
\centering
\includegraphics[width=3.0in]{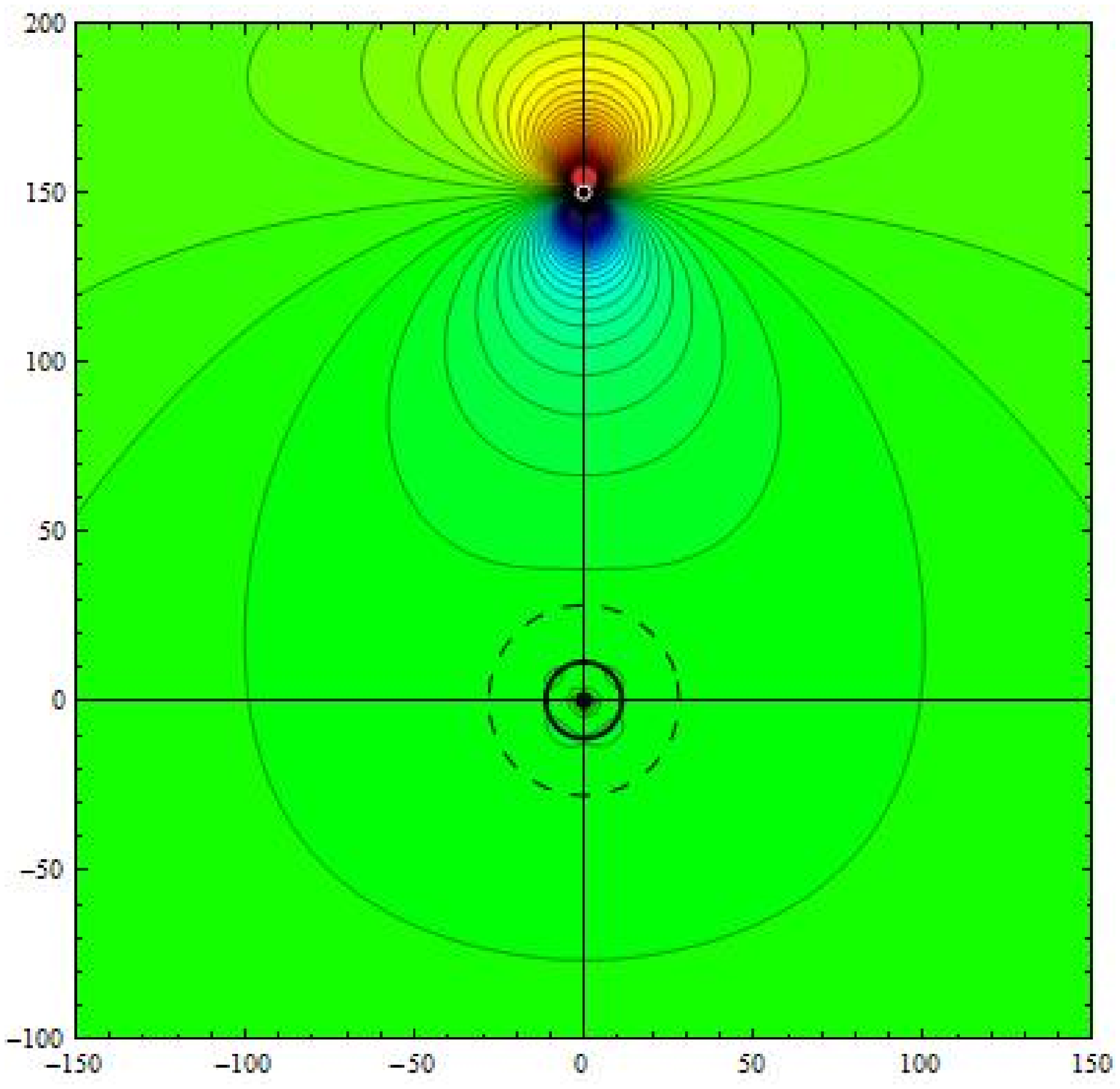}~~~
\includegraphics[width=3.0in]{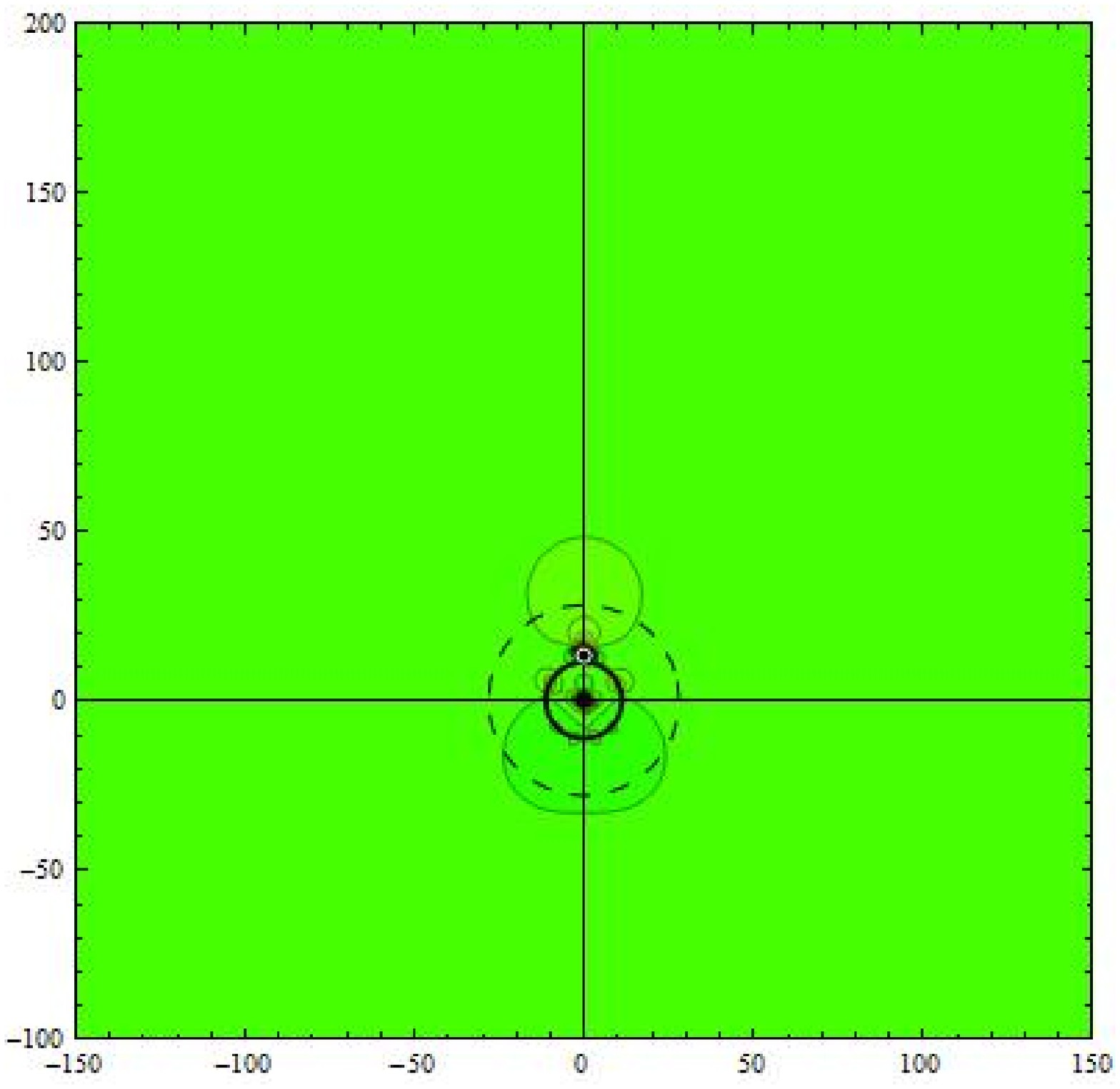}
\begin{minipage}{\textwidth}
\hspace{0.225\textwidth}
(a)
\hspace{0.46\textwidth}
(b)
\end{minipage}
\caption{\label{fig07xx} Contour plot of $|H_\mathrm{z}|$ as a function
of position for a system consisting of a coated cylinder
($r_\mathrm{c} = 1.8$nm, $r_\mathrm{s} = 11.25$nm, $\varepsilon_\mathrm{c} = 1$,
$\varepsilon_\mathrm{s} = -1 + 0.01\,\mathrm{i}$,
$\mu_\mathrm{c} = \mu_\mathrm{s} = 1$)
interacting with a probe cylinder
($a = 2$nm, $\varepsilon = -1 + 0.01\,\mathrm{i}$, $\mu = 1$), and
irradiated by a $H_\mathrm{z}$ polarized
plane wave coming from above. (a): The probe cylinder is outside the cloaking region bounded
by the dashed circle, at a distance of 150nm from the origin.
(b): The probe cylinder is outside the cloaking region at a distance of
13.5nm from the origin. The visibility (\ref{visibility}) takes the values $v=0.526$ (a) and
$v=0.021$ (b). }
\end{figure}

We can examine whether the Ostrich Effect can be reduced by making
appropriate choices of the free parameters of the cloaking system:
$n_\mathrm{c}$, $r_\mathrm{c}$ and $r_\mathrm{s}$. We study the effect on the scattering
cross section of varying these parameters in figures \ref{fig056x} and \ref{fig07x}.
Figure \ref{fig056x}
shows the effect of varying the core index, both below and above the
value of unity used in previous figures.
While the effect of $n_\mathrm{c}$ varies with wavelength, in general one
sees from figure \ref{fig056x} that values around unity deliver the lowest cross
sections. In figure \ref{fig07x} we study the effect on cross section of varying
radii. Here, the cross section values have been normalized by
dividing by the cylinder diameter, to give a dimensionless value.
The geometric parameters $r_\mathrm{c}$ is shown divided by the wavelength,
while the horizontal axis gives the cloaking radius $r_\#$ divided by
$r_\mathrm{s}$. The leftmost curve gives the contour on which the cross
section is equal to 1\% of the geometric value. Along this contour,
if we want to have say $r_\#=2.5 r_\mathrm{s}$, to give a relatively large
cloaked region, we need $r_\mathrm{c}\simeq 0.003 \lambda$ and $r_\mathrm{s}\simeq
6.25 r_\mathrm{c}$, or $r_\mathrm{s}\simeq 0.019 \lambda$. These relatively strict
tolerances illustrate the difficulty of achieving low cross section
values at finite wavelengths. Figures \ref{fig07xx} (a) and (b), show two
field distributions corresponding
to the probe inside and outside the cloaking region, with
$\mathrm{Im}(\varepsilon_{\rm s})$ now set to $0.01$, in order to
make more effective the cloaking action. We can see in these figures that we have achieved a satisfactory
combination of effective cloaking of the probe and virtual elimination of the Ostrich Effect.
We can quantify this by introducing a dimensionless quantity we call {\em visibility}
defined in a similar fashion to the quantity in interference optics:
\begin{equation}
v = \left( \left|H_{\rm z}^{\rm (max)}\right| - \left| H_{\rm z}^{\rm (min)}\right|\right)/
\left( \left|H_{\rm z}^{\rm (max)}\right| + \left| H_{\rm z}^{\rm (min)}\right|\right) .
\label{visibility}
\end{equation}
Here, $|H_{\rm z}^{\rm (min)}|$ and $|H_{\rm z}^{\rm (max)}|$ denote the minima and the
maxima of modulus of $H_{\rm z}$ in the region outside
a circle of radius $r_\#$, if the probe is within the cloaking region, and outside the
minimal circle containing the coated cylinder and the probe, if the probe is outside
the cloaking region. The values of $v$ are $0.526$ for figure \ref{fig07xx}(a) and
$0.021$ for figure \ref{fig07xx}(b), a satisfactorily small value.

\section{Conclusions}

We have presented numerical results displaying clearly the tendency for it to be more difficult
to hide the larger cloaking system than the smaller object it is trying to conceal
from electromagnetic probing, and we have analyzed this effect to quantify to what extent it
can be overcome. Our results are conveniently summarized in figure \ref{fig07xx},
and give size limits on the
cloaking system in terms of the wavelength. These size limits in fact just require both the
cloaking system and the system it is cloaking to be in the quasistatic regime.

We have also shown that this regime in fact sets in at shorter wavelengths for a resonant
cloaking system with a small amount of loss, compared with the case of no loss.
We have confined our studies to cloaking systems which have spatially uniform shells,
but it may
be the case that structured systems of the sort described by Farhat \etal \cite{sasha}
may be designed
which inhibit multipole responses in such a way as to ensure the onset of quasistatic behaviour
at shorter wavelengths than indicated by figure \ref{fig07xx}. This would be valuable in  possibly
simplifying the construction of cloaking systems which operate by plasmonic resonance, while
it would be bringing their geometry closer to that of systems which cloak by refraction.

\ack
Nicolae Nicorovici and Ross McPhedran acknowledge support from the
Australian Research Council's Discovery Projects Scheme. The
latter's work on this project was also supported by the C.N.R.S.,
France.

\end{document}